\documentclass[twocolumn,english,aps,superscriptaddress,pre,longbibliography]{revtex4-1}
\usepackage[T1]{fontenc}
\usepackage[latin9]{inputenc}
\usepackage{color}
\usepackage{babel}
\usepackage{amsmath}
\usepackage{amssymb}
\usepackage{graphicx}
\usepackage{esint}
\usepackage[unicode=true,pdfusetitle,
 bookmarks=true,bookmarksnumbered=false,bookmarksopen=false,
 breaklinks=true,pdfborder={0 0 0},pdfborderstyle={},backref=false,colorlinks=true]
 {hyperref}
\hypersetup{
 linkcolor=blue,citecolor=blue,urlcolor=blue}

\makeatletter
\usepackage{pslatex}

\usepackage{color}
\usepackage{babel}
\usepackage[caption=false]{subfig}

\@ifundefined{showcaptionsetup}{}{%
 \PassOptionsToPackage{caption=false}{subfig}}
\usepackage{subfig}
\makeatother

\begin{document}
\title{Phase transitions in a conservative Game of Life}
\author{André P. Vieira}
\affiliation{Universidade de Sao Paulo, Instituto de Fisica, Rua do Matao, 1371,
05508-090, Sao Paulo, SP, Brazil}
\author{Eric Goles}
\affiliation{Facultad de Ingeniería y Ciencias, Universidad Adolfo Ibáñez, Avenida
Diagonal las Torres 2640, Peñalolén, Santiago, Chile}
\author{Hans J. Herrmann}
\affiliation{Departamento de Física, Universidade Federal do Ceará, 60451-970 Fortaleza,
CE, Brazil}
\affiliation{ESPCI, CNRS UMR 7636 - Laboratoire PMMH, 75005 Paris, France}
\date{\today}
\begin{abstract}
We investigate the dynamics of a conservative version of Conway's
Game of Life, in which a pair consisting of a dead and a living cell
can switch their states following Conway's rules but only by swapping
their positions, irrespective of their mutual distance. Our study
is based on square-lattice simulations as well as a mean-field calculation.
As the density of dead cells is increased, we identify a discontinuous
phase transition between an inactive phase, in which the dynamics
freezes after a finite time, and an active phase, in which the dynamics
persists indefinitely in the thermodynamic limit. Further increasing
the density of dead cells leads the system back to an inactive phase
via a second transition, which is continuous on the square lattice
but discontinuous in the mean-field limit.
\end{abstract}
\maketitle

\section{Introduction}

Since it was proposed by Conway about 50 years ago \citep{Gardner1970},
the cellular automaton known as the Game of Life has been investigated
by statistical physicists as a paradigm for emergent complex behavior
based on simple, local rules. In its original version, cells situated
on a square lattice can be either ``alive'' or ``dead'', and switch
synchronously from one state to the other depending on how many of
their 8 neighboring cells (a Moore neighborhood) are alive. More precisely,
a dead cell becomes alive if exactly 3 of its neighbors are alive,
while a living cell becomes dead unless it has 2 or 3 neighboring
living cells. Dynamical evolution under these rules leads to a variety
of complex behavior, in which living cells are able to exhibit a mixture
of static (or ``still-life''), oscillatory, and progressive (or ``spaceship'')
patterns, depending on the initial conditions. From a computer-science
perspective, these patterns can be used to build a universal Turing
machine \citep{Rendell2002}, which is able to simulate any circuit
and therefore any algorithm. 

From the point of view of statistical physics, the automaton gained
widespread interest due to discussions \citep{Bak1989,Bak1992,Bennett1991,Garcia1993}
on whether it represented an example of self-organized criticality
\citep{Bak1987} in the absence of conserved quantities. It now seems
that the Game of Life is in a slightly subcritical state, corresponding
to a fine-tuned quasicritical nucleation process at the border of
extinction \citep{Alstrom1994,Hemmingsson1996,Blok1997,Reia2014}. 

A related topic is the appearance of phase transitions when some stochastic
ingredient is added. Some previous investigations \citep{Monetti1995,Nordfalk1996}
replaced the original deterministic automaton by a stochastic one,
in which Conway's rules were obeyed or subverted with various probabilities.
By tuning those probabilities, the long-time behavior associated with
the dynamics can be changed from one in which all cells are dead to
another in which life thrives. Both continuous \citep{Nordfalk1996}
and discontinuous \citep{Monetti1995} phase transitions can be observed,
depending on the choice of probabilities. Another possibility is to
keep the deterministic nature of the automaton, but introduce randomness
by changing the nature of the lattice into a small-world network,
obtained by replacing nearest-neighbor links with long-range ones
\citep{Huang2003}. Increasing the probability of such rewiring induces
a continuous nonequilibrium phase transition from an inactive (sparse)
state to an active (dense) one. 

A common feature of these previous studies is that they preserve the
nonconservative nature of the dynamical rules. Here, on the other
hand, we modify the dynamical rules in order to enforce conservation
of the number of cells of both types. We work on the square lattice,
treating the system as asynchronous, and at each time step we randomly
select an ``unsatisfied'' pair of cells, consisting of a living cell
that is to become dead and a dead cell that is to become alive according
to Conway's (local) rules, and we switch their positions, irrespective
of their mutual distance. The evolution freezes if there are no remaining
unsatisfied living or dead cells. By keeping track of the dependence
of the average freezing time and of the average fractions of unsatisfied
cells on the lattice size, we identify a discontinuous nonequilibrium
phase transition, induced by increasing the density of dead cells,
between a state in which the evolution freezes at a finite time and
another state in which the dynamics persists indefinitely. Further
increasing the density of dead cells leads the system back to an inactive
phase via a second but now continuous transition.

This paper is organized as follows. In Sec. \ref{sec:simulations}
we describe the details of our simulations, also discussing some peculiar
finite-size effects, as well as the nature of the phase transitions.
In Sec. \ref{sec:meanfield}, we present a mean-field calculation
which is able to reproduce various features of our simulation results.
We close the paper with a discussion in Sec. \ref{sec:discussion}.

\section{simulations}

\label{sec:simulations}Following Conway \citep{Gardner1970}, we
consider a square lattice and assume that each cell in a $L\times L$
lattice can be either in state $a$ (``alive'') or in state $d$
(``dead''), and has 8 neighboring cells, i.e. a Moore neighborhood.
We implement periodic boundary conditions. A living cell is satisfied
if either 2 or 3 of its neighbors are also alive, and is otherwise
unsatisfied; a dead cell is satisfied unless exactly 3 of its neighbors
are alive. In contrast to the original version of the Game of Life,
we assume a conservative, asynchronous and nonlocal dynamics: at each
time step, we randomly select an unsatisfied living cell and an unsatisfied
dead cell and we switch their states, also checking for changes in
the satisfaction of any neighboring cell, and then we repeat the previous
steps. When there are no more unsatisfied cells in either state $a$
or state $d$, the evolution is frozen, which is bound to happen eventually
for any finite system. Notice that the nonlocal character of the switching
does not imply that the model has a mean-field character, as the satisfaction
rules are still locally defined. This is analogous to what happens
for the conservative contact process \citep{tome2001}, which remains
in the directed-percolation universality class despite nonlocal switches.

\begin{figure}
\begin{centering}
\subfloat[]{\begin{centering}
\includegraphics[width=0.47\columnwidth]{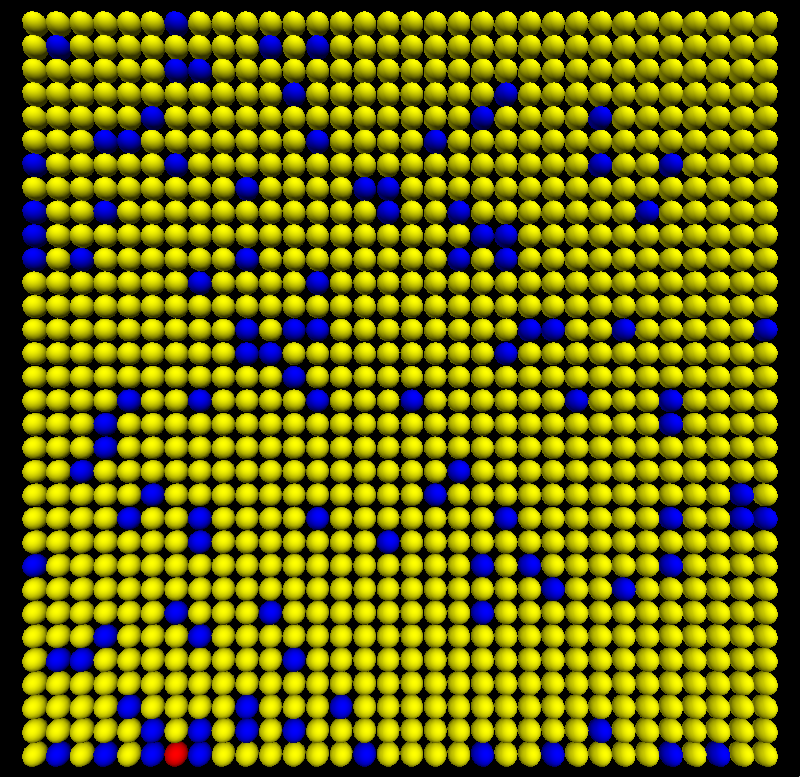}
\par\end{centering}
}\hfill{}\subfloat[]{\begin{centering}
\includegraphics[width=0.47\columnwidth]{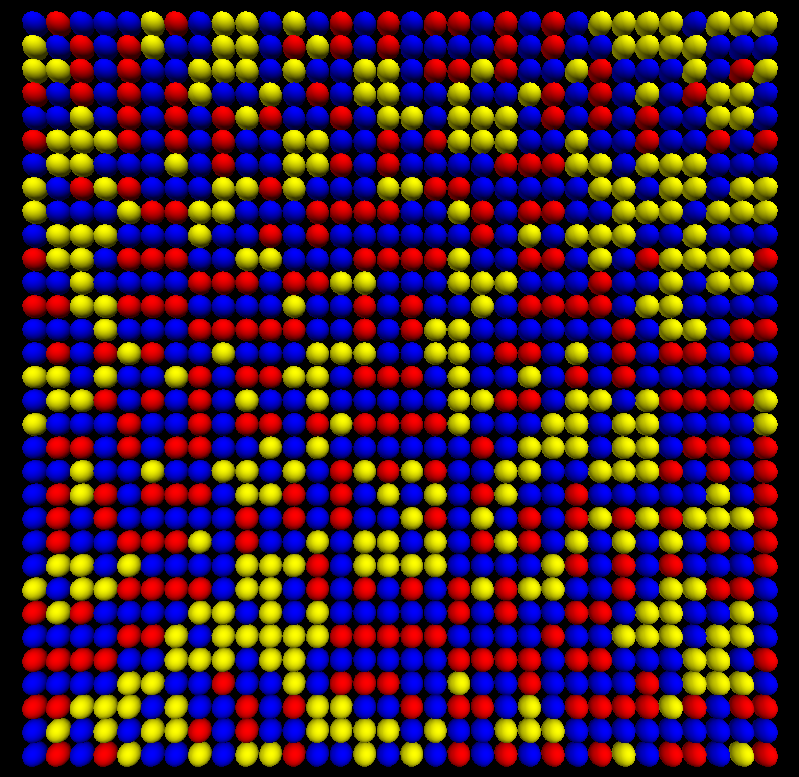}
\par\end{centering}
}
\par\end{centering}
\centering{}\subfloat[]{\begin{centering}
\includegraphics[width=0.47\columnwidth]{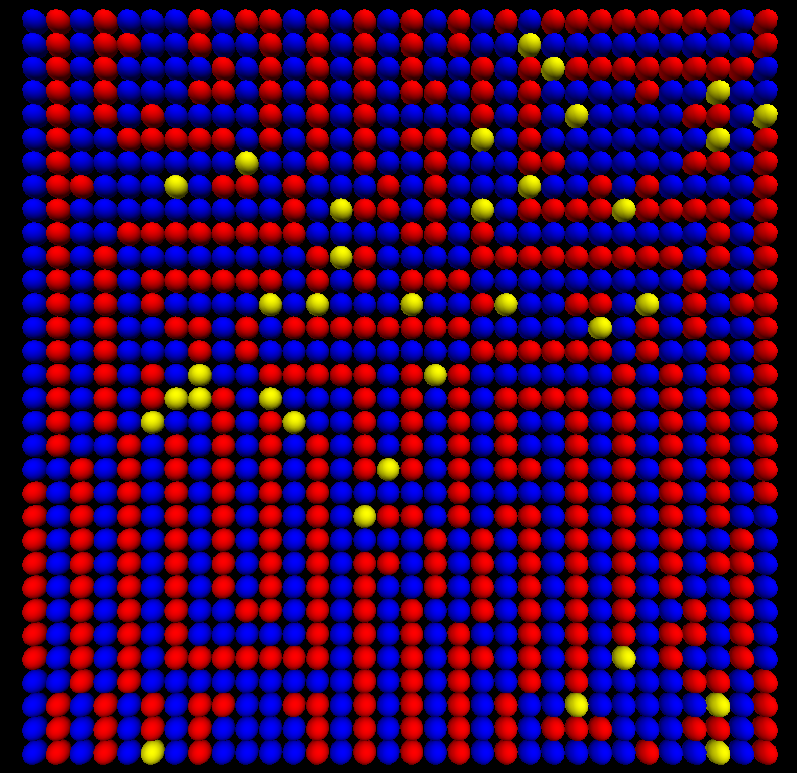}
\par\end{centering}
}\hfill{}\subfloat[]{\begin{centering}
\includegraphics[width=0.47\columnwidth]{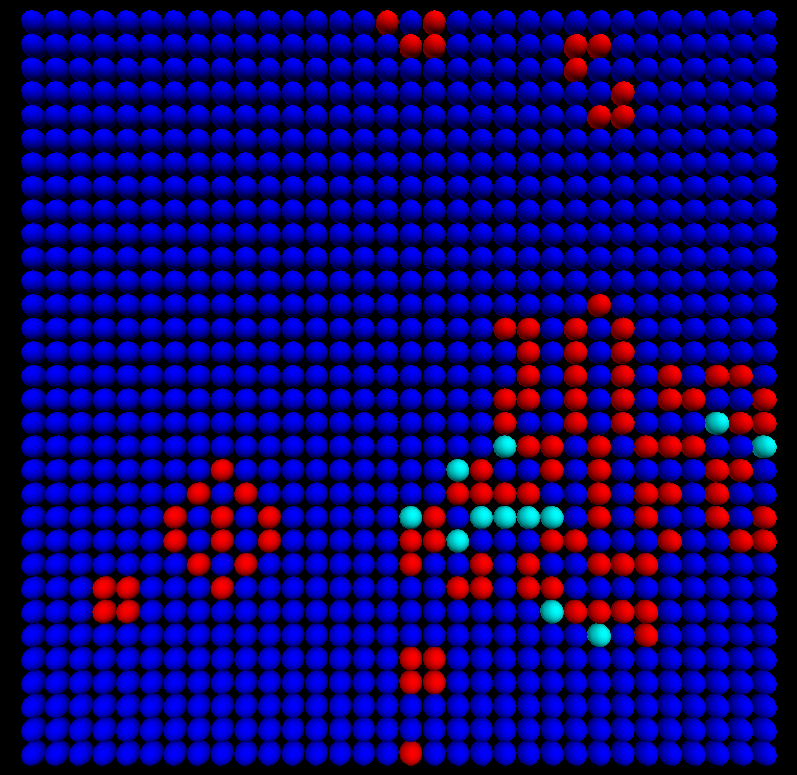}
\par\end{centering}
}\caption{\label{fig:examples}Examples of configurations generated by the conservative
Game of Life upon freezing, on a square lattice with $L=32$. Satisfied
(unsatisfied) living cells are shown in red (yellow), while satisfied
(unsatisfied) dead cells are shown in blue (cyan). The density of
dead cells is (a) $\rho_{d}=0.1$, (b) $\rho_{d}=0.46$, (c) $\rho_{d}=0.54$,
(d) $\rho_{d}=0.9$.}
\end{figure}
Although the conservative and random character of the dynamics does
not allow for the appearance of oscillatory or spaceship patterns,
we do observe familiar static patterns when the density of dead cells
is large enough, as illustrated in Fig. \ref{fig:examples}(d). For
very small densities of dead cells, most living cells are unsatisfied
upon freezing, as shown in Fig. \ref{fig:examples}(a), while for
intermediate densities, dead and living cells arrange themselves in
domains consisting of lines of alternating type, as shown in Figs.
\ref{fig:examples}(b) and (c).

We work with lattice sizes ranging from $L=16$ to $L=23000$, performing
averages over up to $10^{5}$ random initial configurations, and for
each configuration we fix the densities $\rho_{d}$ and $\rho_{a}=1-\rho_{d}$
of cells in states $d$ and $a$, respectively. These densities are
kept invariant by the dynamics. Time increments between consecutive
simulation steps are measured in units of the inverse number of unsatisfied
cells, being given by
\begin{equation}
\Delta t=\frac{1}{N_{u,a}+N_{u,d}},
\end{equation}
in which $N_{u,a}$ and $N_{u,d}$ are the numbers of unsatisfied
cells in states $a$ and $d$, respectively. We keep track of the
fractions of unsatisfied cells in each state, 
\begin{equation}
n_{u,a}\equiv\frac{N_{u,a}}{\rho_{a}L^{2}}\quad\text{and}\quad n_{u,d}\equiv\frac{N_{u,d}}{\rho_{d}L^{2}},
\end{equation}
in which $\rho_{a}L^{2}$ and $\rho_{d}L^{2}$ are respectively the
total number of living and dead cells. Notice from these definitions
that, irrespective of the density $\rho_{d}$ of dead cells, we have
$0\leq n_{u,a}\leq1$ and $0\leq n_{u,d}\leq1$. We also register
the accumulated time $T$ until a simulation freezes, as well as the
survival probability $P_{\text{s}}\left(t\right)$, defined as the
fraction of simulations reaching time $t$.

\begin{figure}
\centering{}\includegraphics[width=0.99\columnwidth]{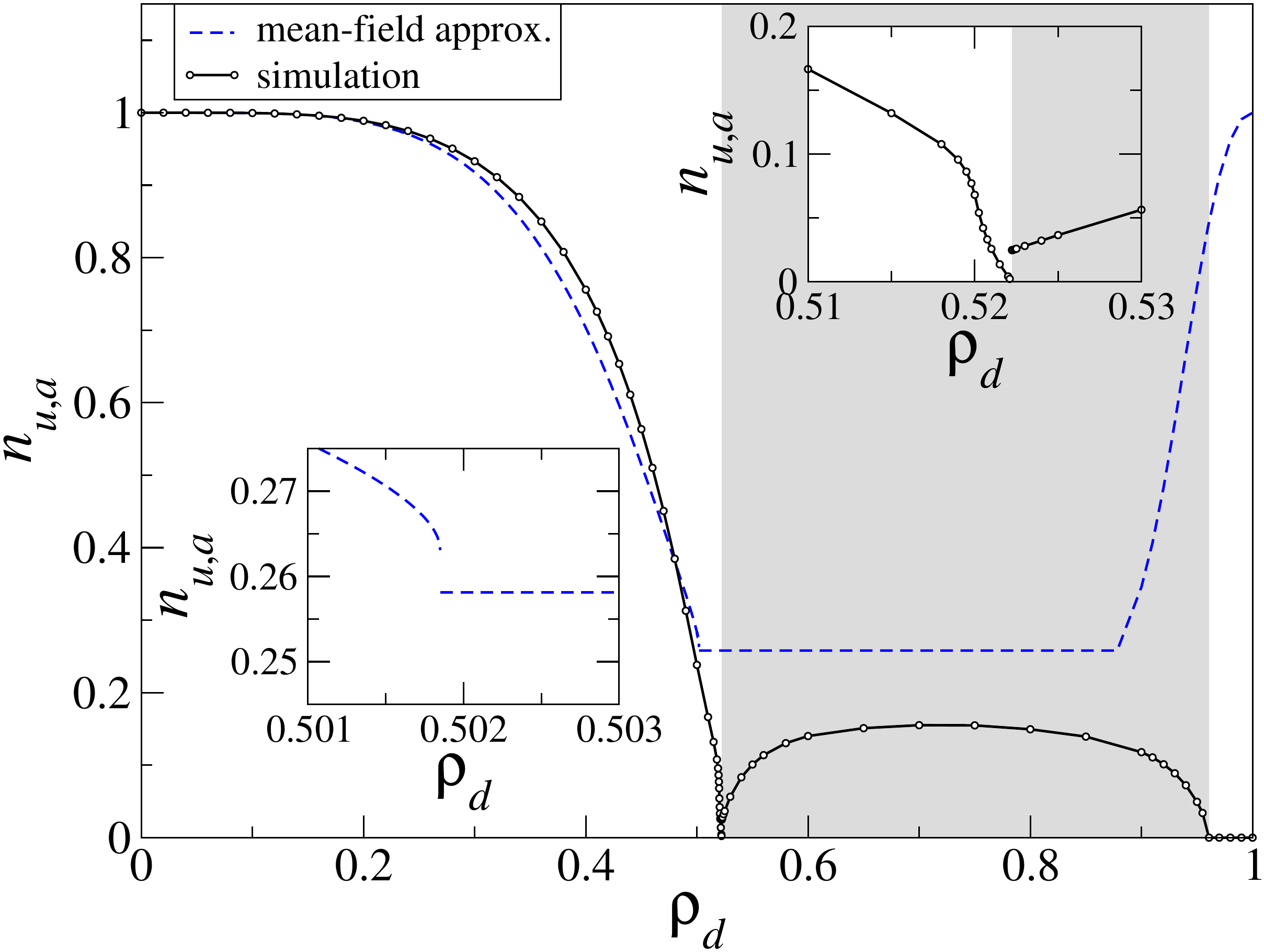}\caption{\label{fig:phase-diagram}Phase diagram of the conservative Game of
Life with a density $\rho_{d}$ of dead cells, as derived from the
average fraction $n_{u,a}$ of unsatisfied living cells. Results from
simulations are shown as black continuous lines with circles, while
mean-field results, discussed in Sec. \ref{sec:meanfield}, are shown
as blue dashed curves. The shaded region corresponds to the active
phase as determined from simulations. Inactive phases exist both in
the small-$\rho_{d}$ and in the large-$\rho_{d}$ limits. The insets
present closer views of the behavior of $n_{u,a}$ around the density
$\rho_{d}^{\text{(1)}}\simeq0.52223$ (simulations) or $\rho_{d}^{(\mathrm{mf})}\simeq0.501850$
(mean-field) separating the small-$\rho_{d}$ inactive phase from
the active phase. Statistical error bars for the simulation curves
are at most the size of the symbols. In the active phase, $n_{u,a}$
is obtained from the $t\rightarrow\infty$ limit of the fraction of
unsatisfied living cells, while in the inactive phases it represents
the value of the same fraction upon freezing, which happens at a finite
time. All simulation values of $n_{u,a}$ correspond to the largest
sizes studied for each value of $\rho_{d}$, and are expected to be
indistinguishable from the infinite-size results at the scale of the
plots. The mean-field approximation predicts an active phase comprising
the region for which $n_{u,a}$ is a constant.}
 
\end{figure}
Depending on the density $\rho_{d}$ of dead cells, we identify three
dynamical regimes, which we now discuss in turn. The resulting phase
diagram is summarized in Fig. \ref{fig:phase-diagram}.

\begin{figure}
\centering{}\subfloat[]{\begin{centering}
\includegraphics[width=0.99\columnwidth]{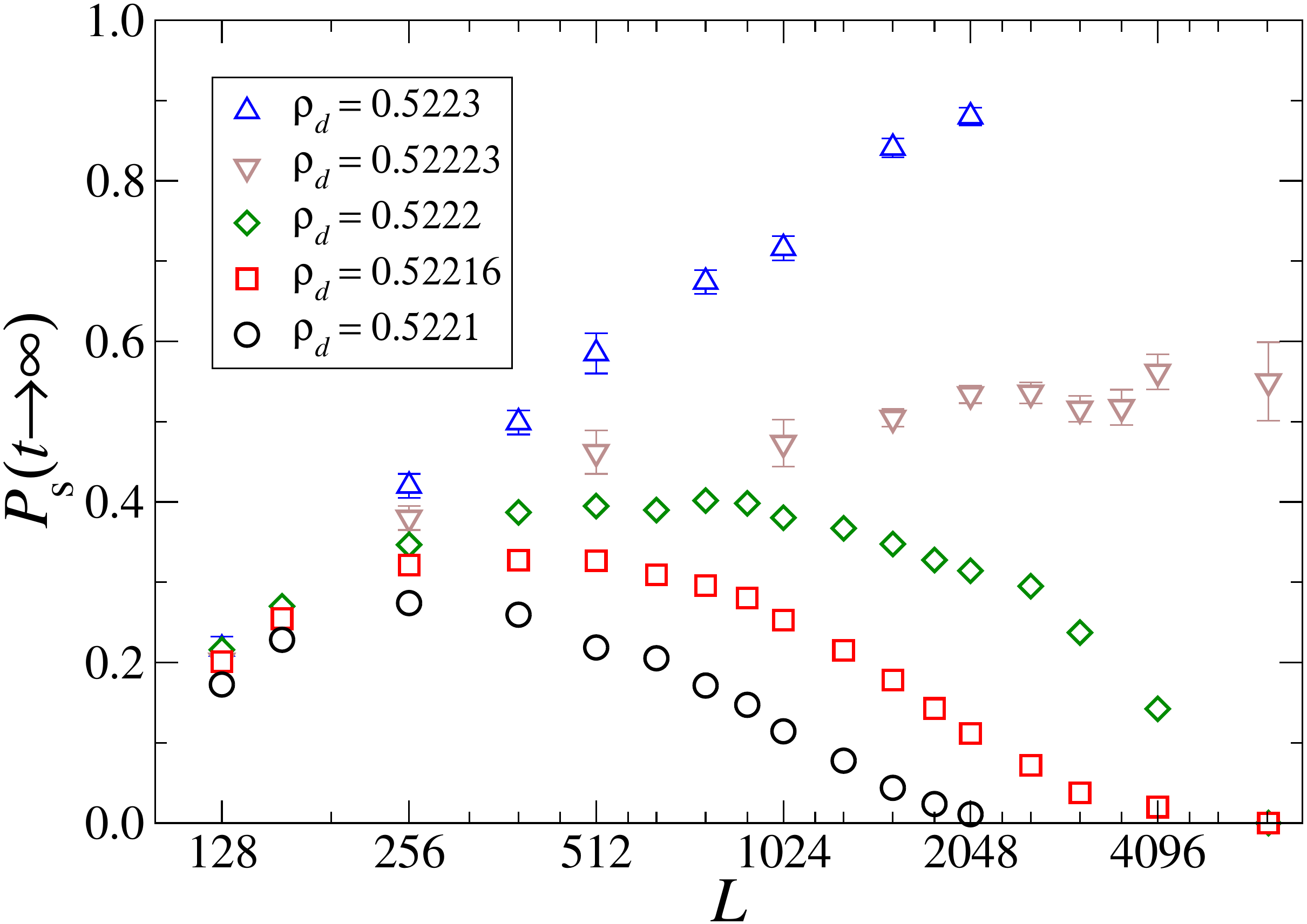}
\par\end{centering}
}\\
\subfloat[]{\begin{centering}
\includegraphics[width=0.99\columnwidth]{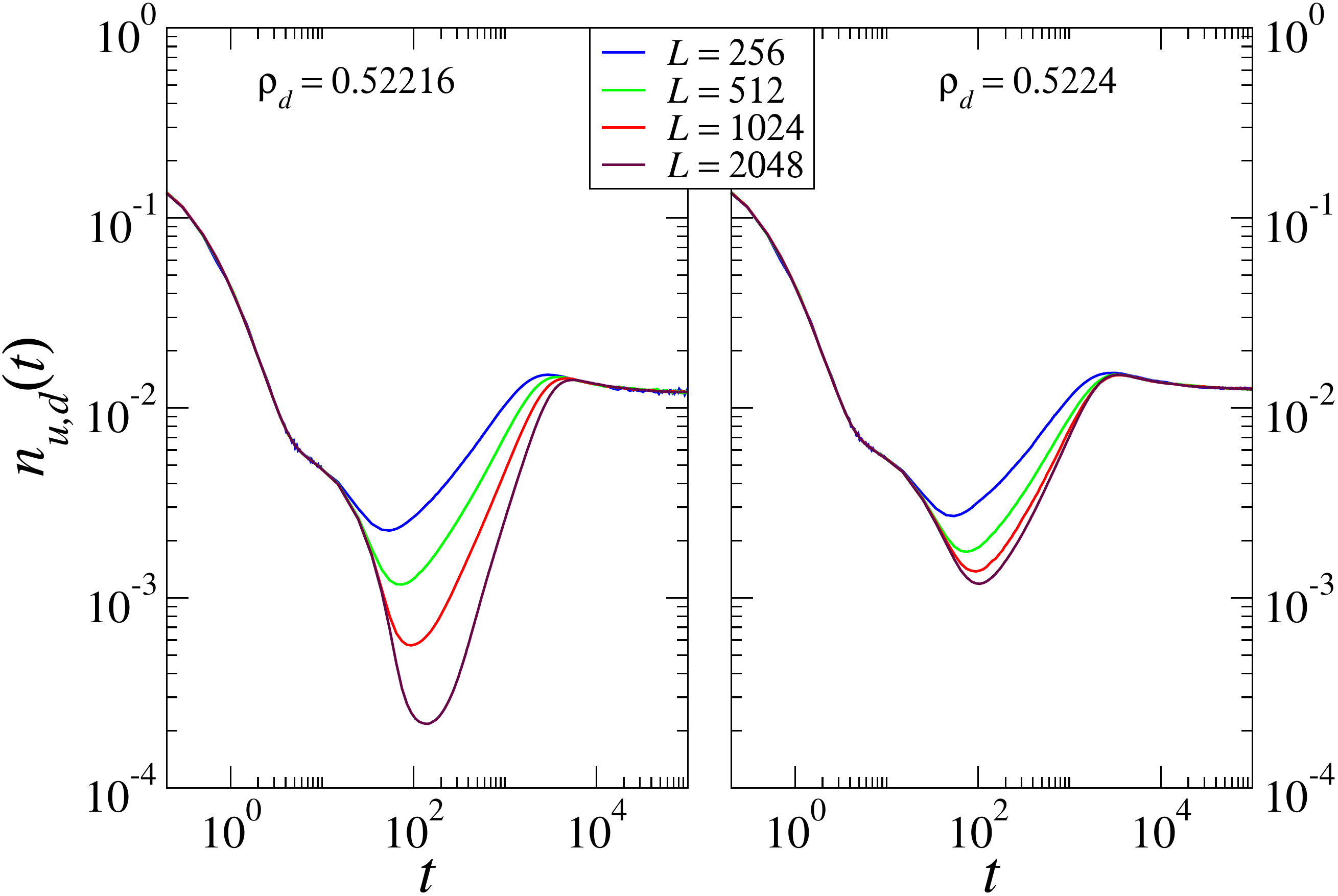}
\par\end{centering}
}\caption{\label{fig:Ninf1vsf_low}(a) Stationary ($t\rightarrow\infty$) survival
probability as a function of the linear system size $L$ for values
of $\rho_{d}$ close to $\rho_{d}^{\left(1\right)}\simeq0.52223$.
Here we take $t=10^{7}$ to mean $t\rightarrow\infty$, but we checked
that taking instead $t=10^{6}$ yields indistinguishable results.
Error bars not shown are at most the size of the symbols. Within statistical
errors, the stationary survival probability exactly for $\rho_{d}=\rho_{d}^{\left(1\right)}$
is asymptotically size-independent, being approximately equal to $50\%$.
(b) Time dependence of the fraction $n_{u,d}$ of unsatisfied dead
cells for $\rho_{d}=0.52216<\text{\ensuremath{\rho_{d}^{\left(1\right)}}}$
and $\rho_{d}=0.5224>\ensuremath{\rho_{d}^{\left(1\right)}}$, illustrating
the marked distinction in the finite-size behavior. For each time
$t$, averages are taken only over simulations reaching $t$.}
\end{figure}
We focus first on the case of sufficiently low densities of dead cells,
$\rho_{d}<\rho_{d}^{\text{(1)}}\simeq0.52223$. As the system size
is increased for a fixed $\rho_{d}\lesssim\rho_{d}^{(1)}$, the long-time
survival probability first increases and then decreases to zero, as
shown in Fig. \ref{fig:Ninf1vsf_low}(a). This is reflected in the
time dependence of the average behavior of the fraction $n_{u,d}$
of unsatisfied dead cells, which, as the system size is increased
for $\rho_{d}=0.52216<\rho_{d}^{\text{(1)}}$, exhibits an exponential
drop visible around $t\simeq100$, whose depth increases with the
system size; see Fig. \ref{fig:Ninf1vsf_low}(b), left panel. We thus
expect that simulations freeze due to the fact that eventually there
are no unsatisfied dead cells, and this is indeed verified. In the
whole regime $\rho_{d}<\rho_{d}^{\text{(1)}}$ we observe that the
average freezing time increases with $\rho_{d}$, diverging approximately
as $-\ln\text{\ensuremath{\left|\rho_{d}-\rho_{d}^{(1)}\right|}}$,
and that the corresponding average fraction $n_{u,a}$ of unsatisfied
living cells decreases with $\rho_{d}$. For this last quantity, this
is shown in Fig. \ref{fig:phase-diagram}. We emphasize that, due
to the fact that, as the survival probability first increases and
then decreases with $L$ for a fixed $\rho_{d}\lesssim\rho_{d}^{(1)}$,
it is necessary to be extra cautious about finite-size effects. 

When $\rho_{d}$ is slightly larger than $\rho_{d}^{\text{(1)}}$,
the drop of $n_{u,d}\left(t\right)$ around $t\simeq100$ becomes
asymptotically independent of $L$, as suggested by Fig. \ref{fig:Ninf1vsf_low}(b),
right panel. On the other hand, the survival probability tends to
$1$ as $L\rightarrow\infty$; see Fig. \ref{fig:Ninf1vsf_low}(a),
upper plot. Notice from Fig. \ref{fig:Ninf1vsf_low}(b) that the $t\rightarrow\infty$
limit of $n_{u,d}\left(t\right)$ is size-independent and varies only
slightly with $\rho_{d}$, although for $\rho_{d}<\rho_{d}^{\text{(1)}}$
that value is never reached for $L\rightarrow\infty$, as the system
freezes at a finite time. We therefore expect a discontinuous behavior
for both $n_{u,d}$ and $n_{u,a}$ at $\rho_{d}=\rho_{d}^{(1)}$,
and this is confirmed in the simulations; see Fig. \ref{fig:phase-diagram},
inset. This discontinuous transition between the small-$\rho_{d}$
inactive phase and the active phase is reproduced by a mean-field
treatment discussed in Sec. \ref{sec:meanfield}.

\begin{figure}
\centering{}\includegraphics[width=0.99\columnwidth]{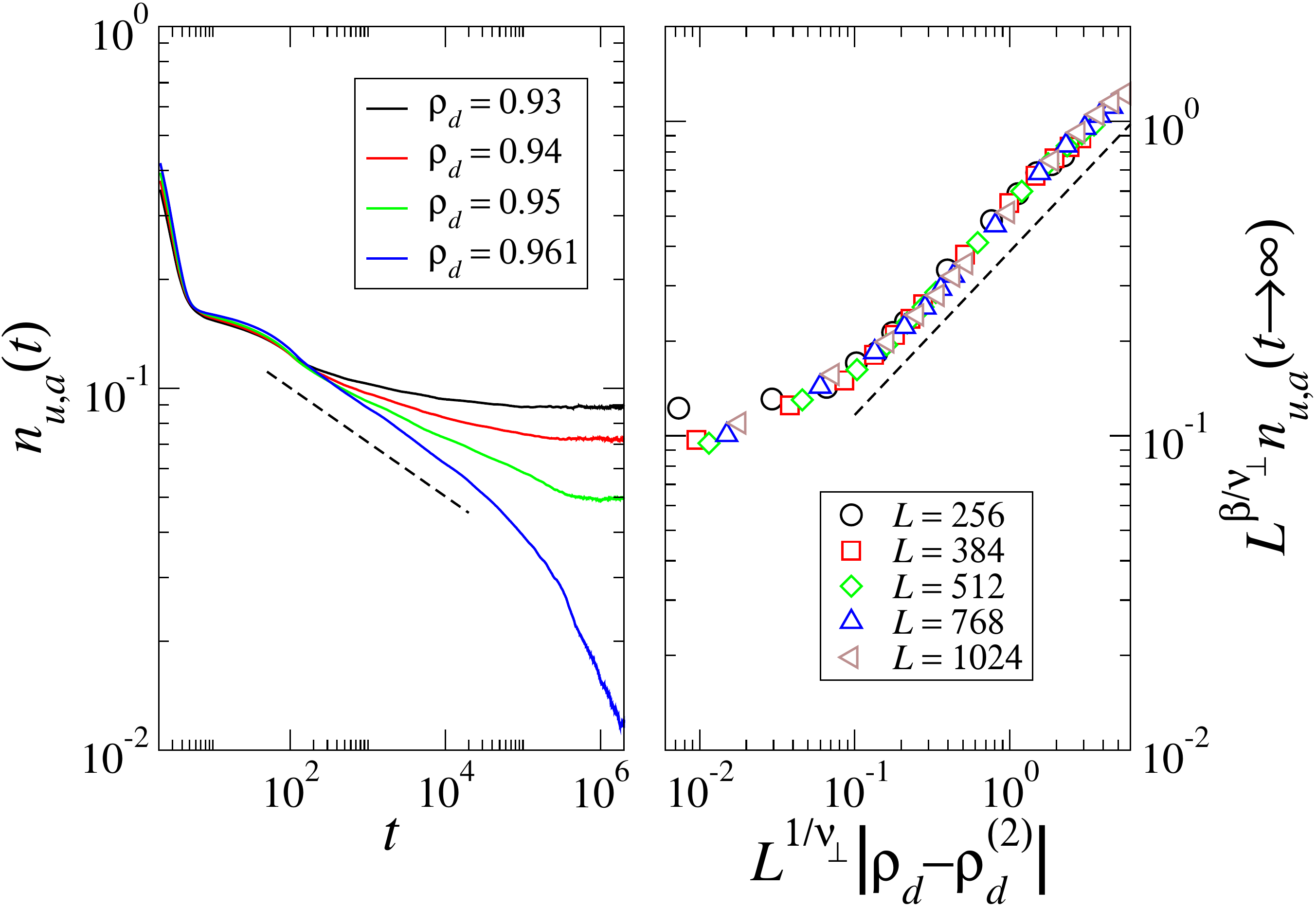}\caption{\label{fig:fss}Left: Time dependence of $n_{u,a}$ for $L=1024$
and large values of $\rho_{d}$ inside the active phase. The dashed
line has slope $-0.15$, the same obtained by fitting data for $\rho_{d}=0.961$
using a power law in the corresponding range. The stronger downward
slope of the curve for $\rho_{d}=0.961$ after $t=10^{4}$ is a finite-size
effect. Right: Finite-size scaling plots of $n_{u,a}\left(t\rightarrow\infty\right)$
for various system sizes $L$, illustrating data collapse with $\rho_{d}^{\left(2\right)}\simeq0.961$,
$\nu_{\perp}\simeq1.54$ and $\beta\simeq0.52$. The dashed line has
slope $\beta$ in log-log scale.}
\end{figure}
Inside the active phase, the freezing time diverges exponentially
with the system size $L$. The long-time value $n_{u,a}\left(t\rightarrow\infty\right)$
first increases with $\rho_{d}$, reaching a maximum around $\rho_{d}\simeq0.7$,
and then decreases, as illustrated in Figs. \ref{fig:phase-diagram}
and \ref{fig:fss} (left panel). When $\rho_{d}$ approaches $\rho_{d}^{\text{(2)}}\simeq0.961$,
the decrease of $n_{u,a}\left(t\rightarrow\infty\right)$ is compatible
with a power law
\[
n_{u,a}\left(t\rightarrow\infty\right)\propto\left|\rho_{d}-\rho_{d}^{(2)}\right|^{\beta},
\]
as suggested by the finite-size scaling analysis presented in Fig.
\ref{fig:fss} (right panel), based on the ansatz
\[
n_{u,a}\left(t\rightarrow\infty\right)=L^{-\beta/\nu_{\perp}}f\left(L^{1/\nu_{\perp}}\epsilon\right),
\]
with $\epsilon=\rho_{d}-\rho_{d}^{(2)}$. The best data collapse,
obtained using the pyfssa package \citep{Sorge2015}, corresponds
to $\rho_{d}^{(2)}=0.961(5)$, $\nu_{\perp}=1.54(5)$ and $\beta\simeq0.52(5)$.
The same parameters yield a collapse of data for $n_{u,d}\left(t\rightarrow\infty\right)$,
although with a narrower scaling region. 

At the critical point, $n_{u,a}\left(t\right)$ follows a power law
$t^{-\theta}$, with $\theta\simeq0.15$; see Fig. \ref{fig:fss}
(left panel). Close to the critical point, and inside the active phase,
we expect this power law to be obeyed up to a relaxation time $\tau_{n}$
which can be estimated by 
\[
\tau_{n}^{-\theta}\propto n_{u,a}\left(t\rightarrow\infty\right)\propto\left|\rho_{d}-\rho_{d}^{(2)}\right|^{\beta},
\]
implying
\[
\tau_{n}\propto\left|\rho_{d}-\rho_{d}^{(2)}\right|^{-\nu_{\parallel}},\quad\nu_{\parallel}=\frac{\beta}{\theta}\simeq3.5.
\]
We therefore expect a dynamical exponent $z=\nu_{\parallel}/\nu_{\perp}$
governing the relation between the relaxation time and the system
size at the critical point,
\begin{equation}
\tau_{n}\propto L^{z},\quad z=\frac{\beta}{\theta\nu_{\perp}}\simeq2.25.\label{eq:tauLz}
\end{equation}

\begin{figure}
\begin{centering}
\subfloat[]{\begin{centering}
\includegraphics[width=0.49\columnwidth]{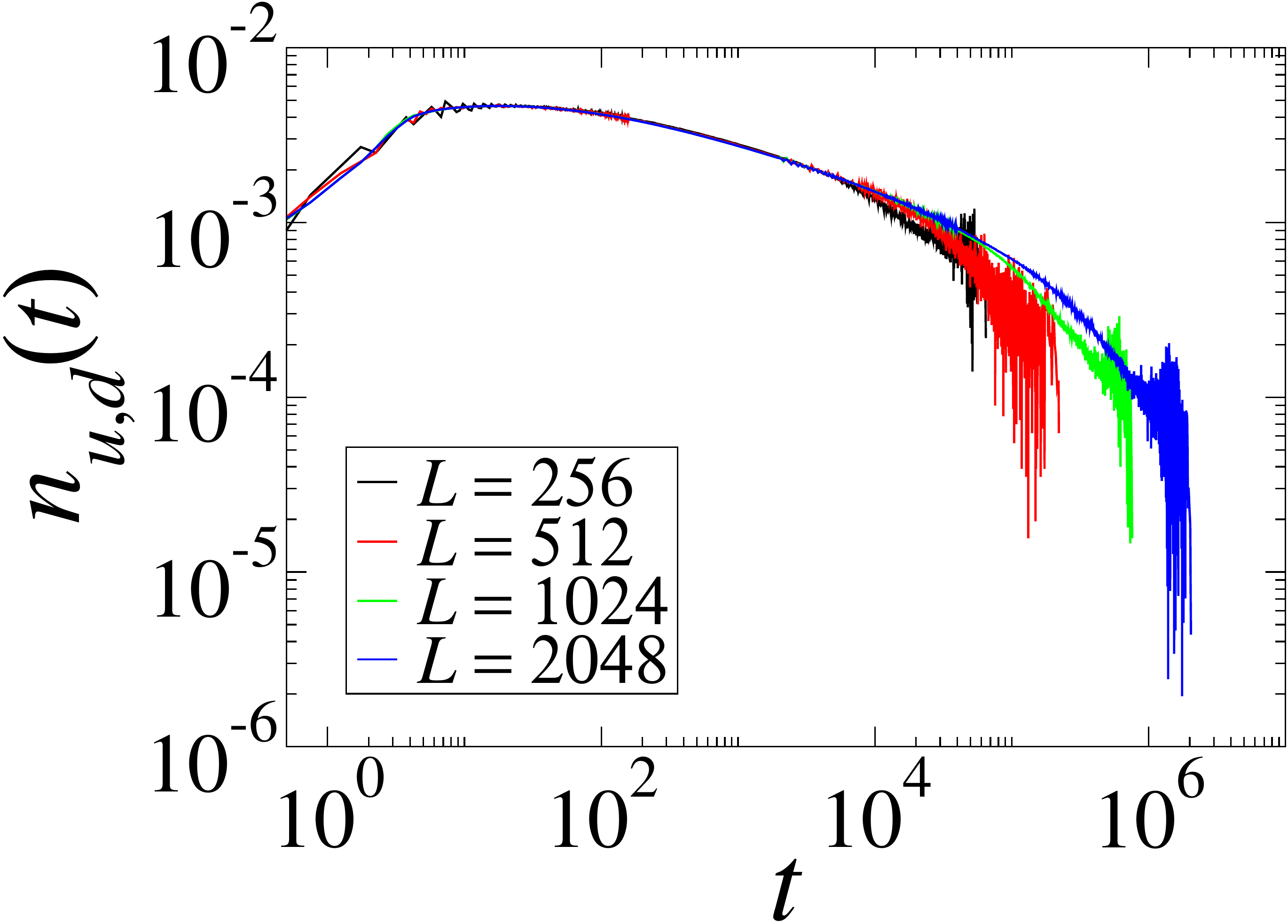}
\par\end{centering}
}\subfloat[]{\begin{centering}
\includegraphics[width=0.49\columnwidth]{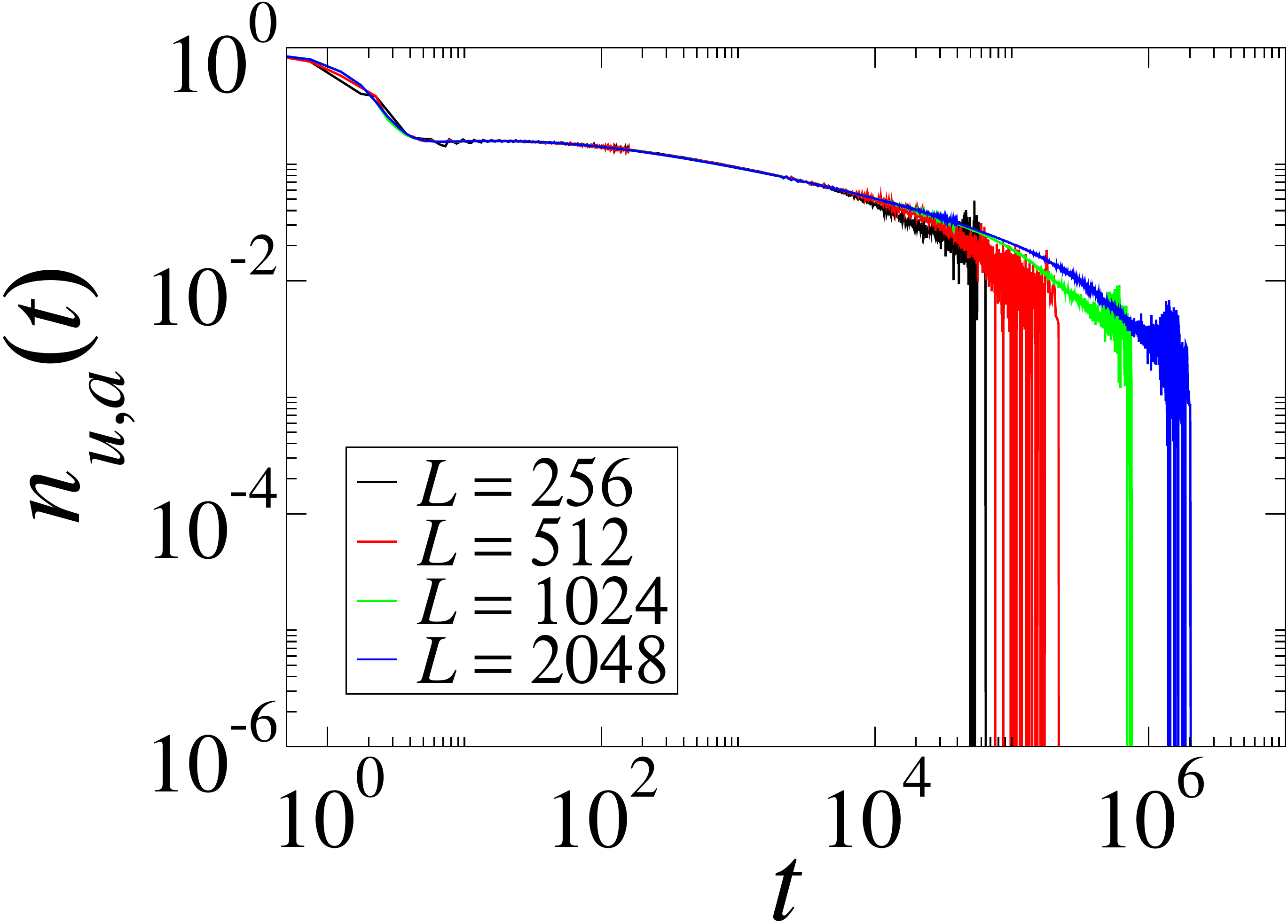}
\par\end{centering}
}
\par\end{centering}
\centering{}\subfloat[]{\begin{centering}
\includegraphics[width=0.49\columnwidth]{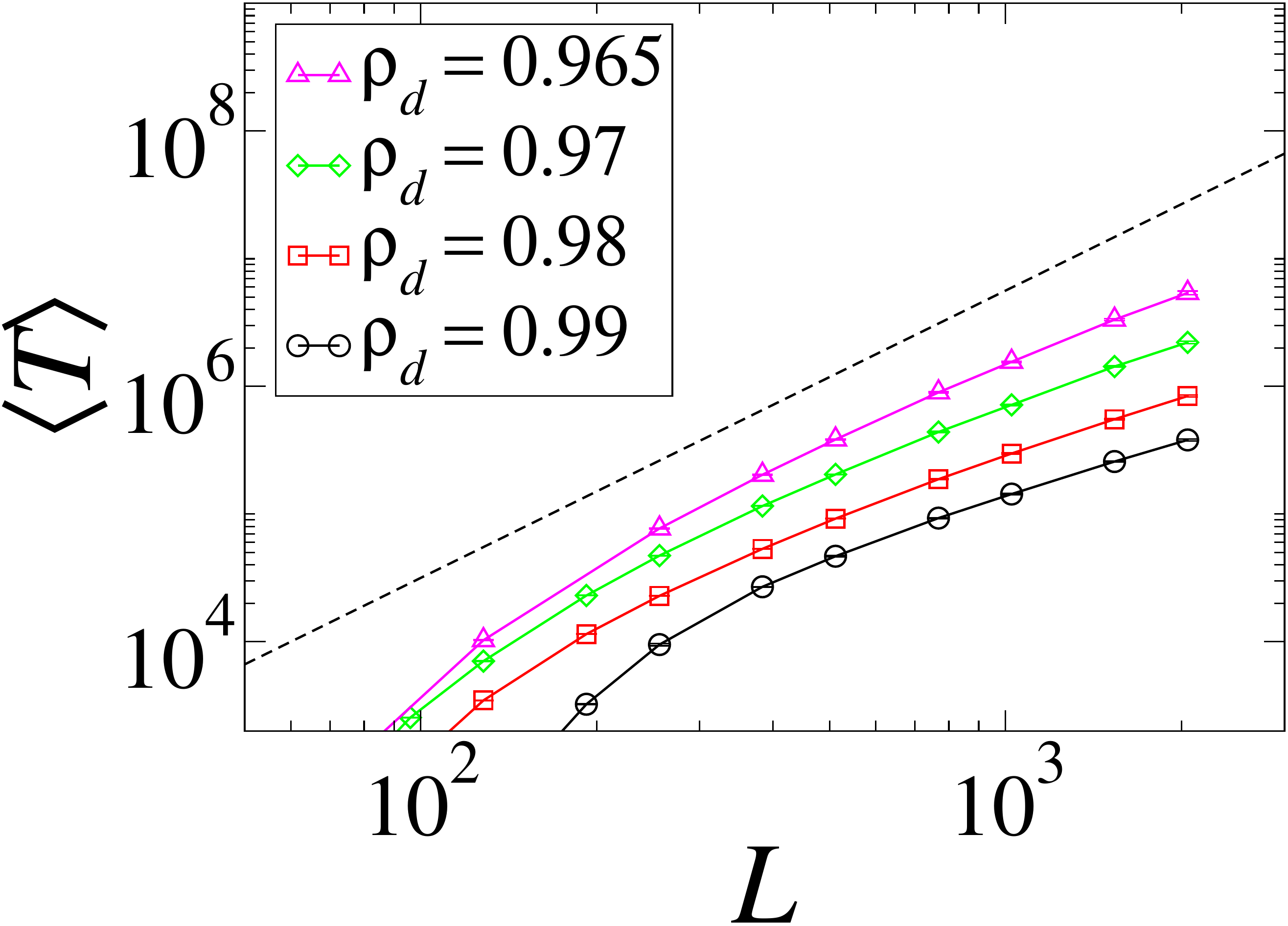}
\par\end{centering}
}\subfloat[]{\begin{centering}
\includegraphics[width=0.49\columnwidth]{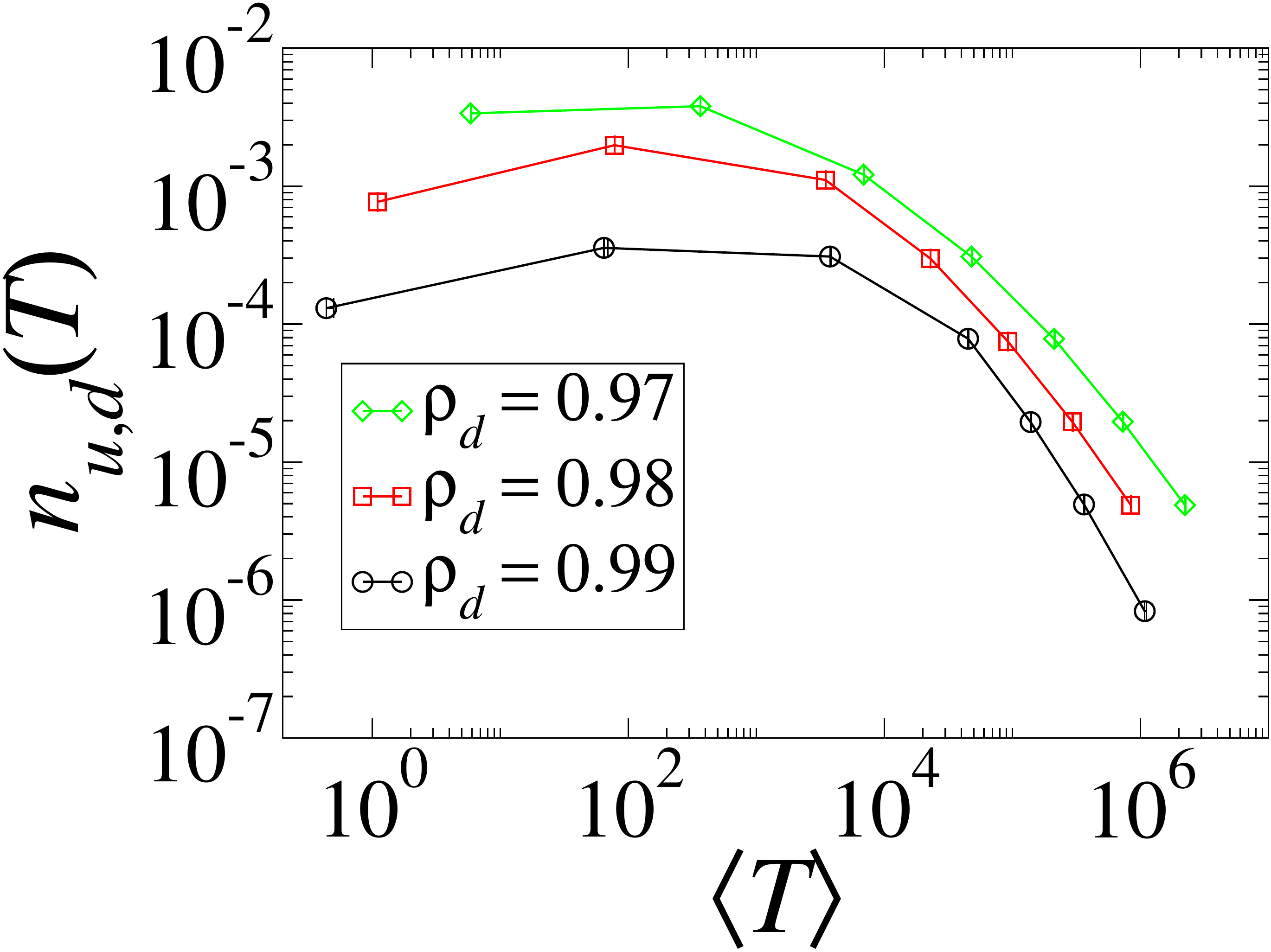}
\par\end{centering}
}\caption{\label{fig:high}Plots of the behavior of various quantities in the
large-$\rho_{d}$ inactive phase. (a) Time dependence of the fraction
of unsatisfied dead cells for $\rho_{d}=0.98$ and various linear
system sizes $L$. (b) Time dependence of the fraction of unsatisfied
living cells for $\rho_{d}=0.98$ and various system sizes. In both
(a) and (b), large fluctuations at later times are associated with
freezing events. (c) Average freezing time as a function of $L$,
for different values of $\rho_{d}$. The dashed line is proportional
to $L^{2.25}$. (d) Parametric plots of the fraction of unsatisfied
dead cells upon freezing, $n_{u,d}\left(T\right)$, versus average
freezing time $\left\langle T\right\rangle $, for different values
of $\rho_{d}$. Each point corresponds to a different choice of $L$
{[}those shown in (c){]}. As clear from (c), $\left\langle T\right\rangle $
increases with $L$.}
\end{figure}
\begin{figure}
\centering{}\includegraphics[width=0.95\columnwidth]{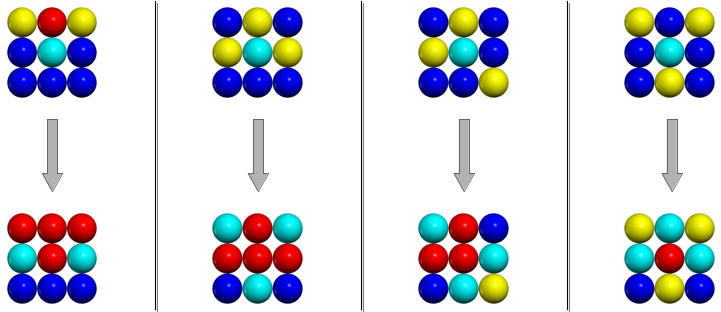}\caption{\label{fig:configs_high}Examples of 4 local configurations around
an initially unsatisfied dead cell, relevant in the large-$\rho_{d}$
limit, for which most likely each configuration is surrounded by dead
cells. Satisfied (unsatisfied) dead cells are shown in blue (cyan),
while satisfied (unsatisfied) living cells are shown in red (yellow).
The upper (lower) row indicates situations before (after) the central
cell is switched to a living cell. Notice the increase in the number
of unsatisfied dead cells from 1 to 2 (first column) or 3 (second,
third and fourth column) and the decrease in the number of unsatisfied
living cells from 3 (first column) or 4 (second, third and fourth
column) to 0 (first and second columns), 1 (third column) or 3 (fourth
column). Take into account that, although not shown in the pictures,
the central dead cells in the upper row become satisfied after switching,
while the central living cells in the lower row were unsatisfied before
switching. }
\end{figure}
Finally, we discuss the case of even larger densities of dead cells,
$\rho_{d}>\rho_{d}^{(2)}$. Now the fraction of living cells is so
small that starting from a random configuration most living cells
only have dead neighbors, so that the fraction of unsatisfied living
cells is very close to unity, while the fraction of unsatisfied dead
cells is very small. As shown in Figs. \ref{fig:high}(a) and (b),
$n_{u,d}\left(t\right)$ first increases, then reaches a maximum before
starting to decrease with time, whereas $n_{u,a}\left(t\right)$ steeply
decreases as $n_{u,d}\left(t\right)$ increases, and then closely
follows the behavior of $n_{u,d}\left(t\right)$ at later times. These
long-time decays are faster than a power law, but slower than an exponential,
possibly suggesting a power law with logarithmic corrections. The
initial behavior of $n_{u,d}\left(t\right)$ and $n_{u,a}\left(t\right)$
can be understood by inspection of the relevant local configurations
around an unsatisfied dead cell to be switched. Due to the small number
of living cells, these relevant configurations are those in which,
before the switch, almost all neighboring living cells are unsatisfied,
while all neighboring dead cells are satisfied. This changes after
the switch, as illustrated for some configurations in Fig. \ref{fig:configs_high}.
As time advances, switches modify the distribution of configurations,
interrupting and then reversing the increase of $n_{u,d}\left(t\right)$.
For sufficiently large $L$, the dynamics is eventually frozen due
to the disappearance of unsatisfied living cells, in contrast to what
happens for small $\rho_{d}$, a case in which freezing is due to
the disappearance of unsatisfied dead cells. Nevertheless, both fractions
of unsatisfied cells upon freezing approach zero in the thermodynamic
limit, as shown in Figs. \ref{fig:high}(a) and (b). 

On the other hand, the average freezing time $\left\langle T\right\rangle $
grows with the system size $L$ at most with a power law rather than
an exponential form for large $L$; see Fig. \ref{fig:high}(c). Assuming
a linear relation between $\left\langle T\right\rangle $ and $\tau_{n}$
at the critical point, that power law would be $L^{z}$, with $z\simeq2.25$,
which is compatible with the behavior observed in Fig \ref{fig:high}(c).
We also observe that the distribution of freezing times (not shown)
is quite narrow, with the survival probability remaining equal to
unity up to approximately the average freezing time for each choice
of $\rho_{d}$ and $L$, after which it decreases to zero exponentially,
with a characteristic time roughly equal to the average freezing time
itself. This is therefore an inactive phase, but with a different
character than the one observed for small values of $\rho_{d}$, in
which characteristic times remain finite in the $L\rightarrow\infty$
limit. 

Finite-size effects in the large-$\rho_{d}$ inactive phase are influenced
by the probability of a dead cell being unsatisfied in the initial
random configuration. The system size $L_{d}$ for which we expect
one unsatisfied dead cell in the initial configuration, therefore
kicking off the dynamics for most configurations, is given by the
solution of
\begin{equation}
\frac{1}{\rho_{d}L_{d}^{2}}=\left(\begin{array}{c}
8\\
3
\end{array}\right)\rho_{d}^{5}\rho_{a}^{3},\label{eq:nudhigh}
\end{equation}
in which we imposed the condition that a dead cell is unsatisfied
only if it has exactly 3 living cells among its 8 neighbors. The distinction
between the regimes $L<L_{d}$ and $L>L_{d}$ is reflected on the
behavior of $n_{u,d}\left(T\right)$, the fraction of unsatisfied
dead cells upon freezing, versus the average freezing time, shown
in Fig. \ref{fig:high}(d) for three values of $\rho_{d}$ in the
large-$\rho_{d}$ inactive phase. When $L\ll L_{d}$, $n_{u,d}\left(T\right)$
takes values of the order of the right-hand side of Eq. (\ref{eq:nudhigh}),
whereas it becomes smaller and smaller as $L$ (and therefore $\left\langle T\right\rangle $)
increases, leading to the interpretation that this corresponds to
an inactive phase with somewhat peculiar properties, resembling a
critical phase.

The set of critical exponents for the transition between the active
phase and the large-$\rho_{d}$ inactive phase is therefore $\beta\simeq0.52$,
$\nu_{\perp}\simeq1.54$, $\nu_{\parallel}\simeq3.5$, $z\simeq2.25$
and $\theta\simeq0.15$. This set is quite distinct both from the
ones characterizing the directed-percolation universality class and
from the sets obtained for conservative (fixed-energy) sandpile models
\citep{Vespignani2000}, as well as the sets corresponding to other
(nonconservative) variants of the Game of Life \citep{Nordfalk1996,Huang2003}.
It is also quite different from the continuous nonequilibrium phase
transitions mean-field exponents $\beta=\theta=\nu_{\parallel}=1$,
$\nu_{\perp}=1/2$ and $z=2$ \citep{Marro2005,Tome2015}.

\section{A mean-field approximation}

\label{sec:meanfield}Features resembling the ones described in the
previous Section can be reproduced by a simple mean-field calculation,
which can provide an estimate of the dependence, on the density of
dead cells, of the asymptotic values of $n_{u,a}$ and $n_{u,d}$.
The calculation proceeds by disregarding spatial correlations, effectively
replacing the Moore-neighborhood square lattice by a Cayley tree with
coordination number 8, which has no loops. (As a matter of fact, as
we are interested in what happens in the deep interior of the tree,
we effectively work on the Bethe lattice; see e.g. Ref. \citep{Stauffer2003}.) 

The relevant variables for the calculation are the average fractions
of cells of both types having $k$ living cells as their neighbors.
We denote these fractions by $\phi_{d,k}$ for the dead cells and
by $\phi_{a,k}$ for the living cells. We have 
\begin{equation}
n_{u,d}=\phi_{d,3}\quad\text{and}\quad n_{u,a}=1-\phi_{a,2}-\phi_{a,3}.
\end{equation}
We now consider all the possible exchanges that can happen in the
lattice, which involve switching the positions of a selected unsatisfied
dead cell (that always has exactly 3 neighboring living cells) and
of a selected unsatisfied living cell (a cell with less than 2 or
more than 3 neighboring living cells). We ignore the possibility that
the selected unsatisfied cells are mutual neighbors, a situation which
occurs with negligible probability in the thermodynamic limit. 

By analyzing each possible exchange at a time, we can keep track of
the change in the average number of cells of each type having $k$
neighboring living cells ($0\leq k\leq8$). Denoting these numbers
by $N_{a,k}$ and $N_{d,k}$, with
\begin{equation}
N_{a,k}=\rho_{a}L^{2}\phi_{a,k}\quad\text{and}\quad N_{d,k}=\rho_{d}L^{2}\phi_{d,k},\label{eq:nf}
\end{equation}
in which here $L^{2}$ denotes the number of sites in the lattice,
the corresponding changes can be calculated by taking into account
that, given a specific movement involving a selected unsatisfied living
cell with $\ell$ living neighbors: (i) the selected unsatisfied dead
cell moves into a neighborhood which contains $\ell$ living cells,
thus reducing $N_{d,3}$ and increasing $N_{d,\ell}$ both by $p_{\ell}$,
the probability of selecting an unsatisfied living cell with $\ell$
living neighbors; (ii) the selected unsatisfied living cell moves
into a neighborhood containing 3 living cells, thus increasing $N_{a,3}$
and reducing $N_{a,\ell}$ both by $p_{\ell}$; (iii) the old neighbors
of the selected dead cell now have one more living neighbor, reducing
$N_{d,k}$ and $N_{a,k}$ while increasing $N_{d,k+1}$ and $N_{a,k+1}$
by amounts which depend both on $\ell$ and on the probabilities of
finding a neighbor with $k$ living neighbors ($0\leq k\leq7$); (iv)
the old neighbors of the selected living cell now have one less living
neighbor, reducing $N_{d,k}$ and $N_{a,k}$ while increasing $N_{d,k-1}$
and $N_{a,k-1}$ by amounts which depend both on $\ell$ and on the
probabilities of finding a neighbor with $k$ living neighbors ($1\leq k\leq8$).
In order to properly account for the various possibilities and their
respective probabilities, it should be kept in mind that $\phi_{a,k}$
and $\phi_{d,k}$, for a given $k$, combine different local configurations
containing $k$ neighboring living cells. 

Going through the above considerations, we can write
\begin{multline}
{\displaystyle \begin{array}{ccc}
\Delta N_{s,k} & = & {\displaystyle \sum_{\ell,k}}\left[U_{s,k,\ell}+W_{s,k,\ell}\phi_{s,k-1}^{\left(8\right)}+X_{s,k,\ell}{\displaystyle \phi_{s,k}^{\left(8\right)}}\right.\\
 &  & +\left.Y_{s,k,\ell}\phi_{s,k}^{\left(0\right)}+Z_{s,k,\ell}\phi_{s,k+1}^{\left(0\right)}\right]p_{\ell},
\end{array}}\label{eq:Delta_N}
\end{multline}
in which $s\in\left\{ a,d\right\} $ indicates the type of cell, $U_{s,k,\ell}\in\left\{ 0,\pm1\right\} $
comes from points (i) and (ii) above, $W_{s,k,\ell}\geq0$ and $X_{s,k,\ell}\leq0$
come from the increases and decreases in $N_{s,k}$ due to point (iii)
above, while $Y_{s,k,\ell}\leq0$ and $Z_{s,k,\ell}\geq0$ come from
the decreases and increases in $N_{s,k}$ due to point (iv) above.
Explicitly, we have $U_{s,k,\ell}=W_{s,k,\ell}=X_{s,k,\ell}=Y_{s,k,\ell}=Z_{s,k,\ell}=0$
for $\ell=2$ or $\ell=3$, while, for $\ell\neq2$ and $\ell\neq$3,
\begin{equation}
U_{d,k,\ell}=\delta_{k,\ell}\ \left(k\neq3\right),\quad U_{d,3,\ell}=-1,
\end{equation}
\begin{equation}
U_{a,k,\ell}=-\delta_{k,\ell}\ \left(k\neq3\right),\quad U_{a,3,\ell}=1,
\end{equation}
\begin{equation}
W_{d,k,\ell}=5\left(1-\delta_{k,0}\right),\quad W_{a,k,\ell}=3\left(1-\delta_{k,0}\right),
\end{equation}
\begin{equation}
X_{d,k,\ell}=-5\left(1-\delta_{k,8}\right),\quad X_{a,k,\ell}=-3\left(1-\delta_{k,8}\right),
\end{equation}
\begin{equation}
Y_{d,k,\ell}=-\text{(}8-k)\left(1-\delta_{k,0}\right),\quad Y_{a,k,\ell}=-k\left(1-\delta_{k,0}\right),
\end{equation}
\begin{equation}
Z_{d,k,\ell}=(8-k)\left(1-\delta_{k,8}\right),\quad Z_{a,k,\ell}=k\left(1-\delta_{k,8}\right).
\end{equation}

The factors 
\[
\phi_{s,k}^{\left(8\right)}\equiv\frac{C_{k}^{\left(8\right)}\phi_{s,k}}{\sum_{j=0}^{7}C_{j}^{\left(8\right)}\phi_{s,j}}
\]
in Eq. (\ref{eq:Delta_N}), with
\[
C_{k}^{\left(8\right)}\equiv\binom{7}{k}/\binom{8}{k},
\]
represent the conditional probabilities that a type-$s$ old neighbor
of the selected dead cell had $k$ living neighbors, given the allowed
range $0\leq k\leq7$, while the analogous factors
\[
\phi_{s,k}^{\left(0\right)}\equiv\frac{C_{k}^{\left(0\right)}\phi_{s,k}}{\sum_{j=1}^{8}C_{j}^{\left(0\right)}\phi_{s,j}},
\]
with
\[
C_{k}^{\left(0\right)}\equiv\binom{7}{k-1}/\binom{8}{k},
\]
represent the conditional probabilities that a type-$s$ old neighbor
of the selected living cell had $k$ living neighbors, given the allowed
range $1\leq k\leq8$. The binomial coefficients account for the number
of ways of arranging the 7 remaining neighbors of a site neighboring
a central site, out of the various possibilities for the configurations
of those neighbors. Finally, $p_{\ell}$ can be written as
\[
p_{\ell}=\frac{\phi_{1,\ell}}{1-\phi_{1,2}-\phi_{1,3}}\ \left(\ell\neq2,3\right),
\]
with $p_{2}=p_{3}=0$. 

We can write differential equations for the time evolution of the
fractions $\phi_{s,k}$ by noting that, as stated in the previous
Section, a single exchange advances time by $\Delta t$ such that
\begin{equation}
\left(\Delta t\right)^{-1}=N_{u,d}+N_{u,a}=L^{2}\left[\rho_{d}\phi_{0,3}+\rho_{a}\left(1-\phi_{1,2}-\phi_{1,3}\right)\right].
\end{equation}
In the limit $L\rightarrow\infty$, $\Delta t$ approaches zero and
we obtain, by using Eq. (\ref{eq:nf}),
\begin{equation}
\frac{d\phi_{s,k}}{dt}=\frac{\left[\rho_{d}\phi_{0,3}+\left(1-\rho_{d}\right)\left(1-\phi_{1,2}-\phi_{1,3}\right)\right]}{\rho_{s}}\Delta N_{s,k},\label{eq:dfd}
\end{equation}
with $\Delta N_{s,k}$ given by Eq. (\ref{eq:Delta_N}). 

\begin{figure}
\begin{centering}
\includegraphics[width=0.95\columnwidth]{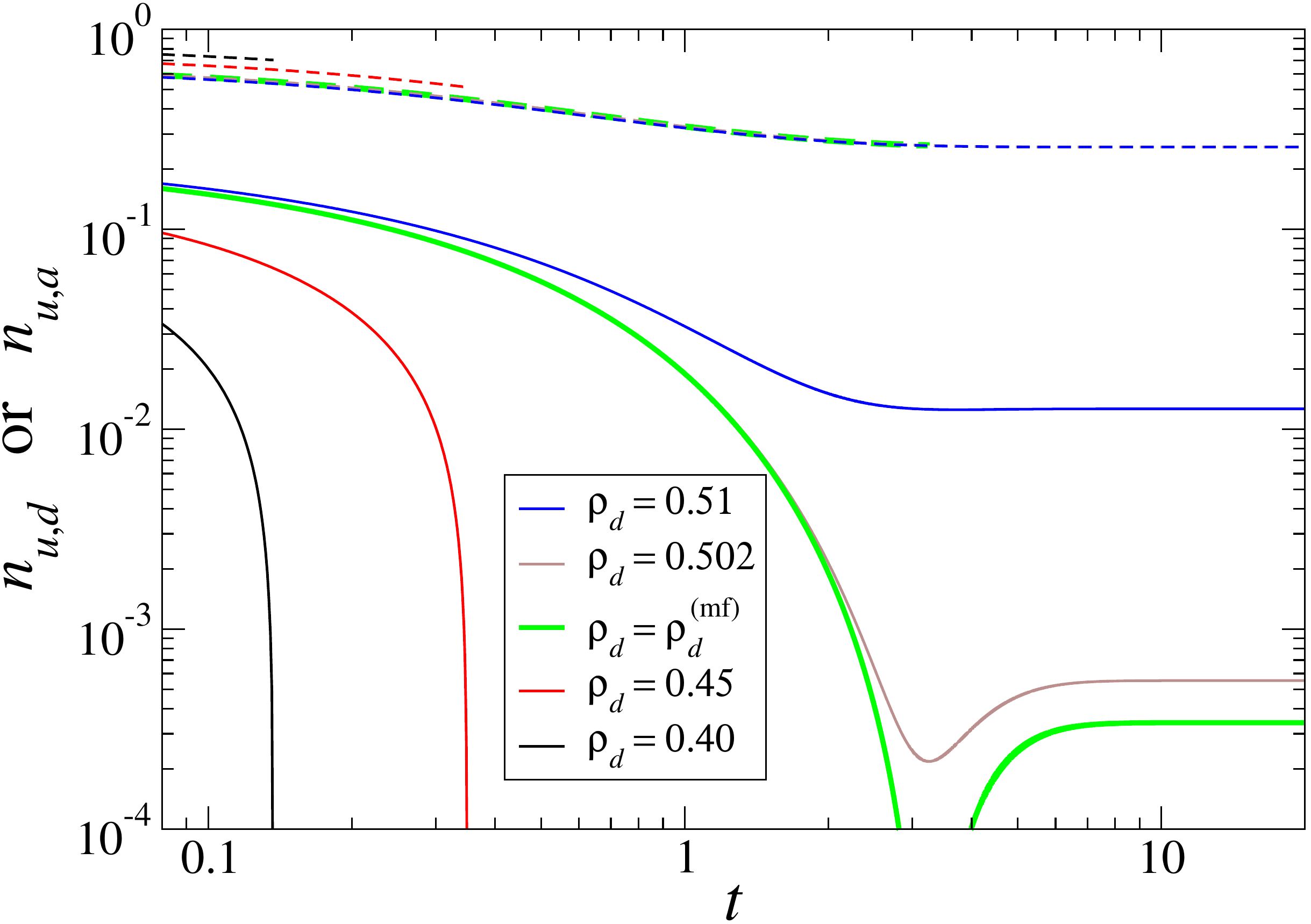}
\par\end{centering}
\caption{\label{fig:Ninf0vst_mf}Time dependence of the fraction of unsatisfied
dead cells $n_{u,d}$ (solid curves) and of unsatisfied living cells
$n_{u,a}$ (dashed curves with the same corresponding color) according
to the mean-field approximation, for various values of the density
$\rho_{d}$ of dead cells. Notice that $n_{u,d}$ does not reach zero
at a finite time for the largest densities shown (brown and blue solid
curves), $\rho_{d}=0.502>\rho_{d}^{\text{(mf)}}$ and $\rho_{d}=0.51>\rho_{d}^{\text{(mf)}}$,
while for $\rho_{d}=\rho_{d}^{\text{(mf)}}$ (thick green curve) $n_{u,d}$
becomes infinitesimally close to zero (hidden by the log vertical
scale) before rising again and asymptotically approaching a value
$3.4\times10^{-4}$ as $t\rightarrow\infty$. The dashed curves for
$\rho_{d}=\rho_{d}^{\text{(mf)}}$, $\rho_{d}=0.502$ and $\rho_{d}=0.51$
are almost indistinguishable at this scale.}
\end{figure}
We numerically solved the differential equations (\ref{eq:dfd}) starting
from random initial conditions, which correspond to
\[
\phi_{s,k}\left(0\right)=\left(\begin{array}{c}
8\\
k
\end{array}\right)\rho_{d}^{8-k}\rho_{a}^{k}.
\]
For $0\leq\rho_{d}\leq\rho_{d}^{\text{(mf)}}\simeq0.501850$, $\phi_{d,3}$
(equal to the fraction of unsatisfied dead cells $n_{u,d}$) is the
only $\phi_{s,k}$ that eventually reaches zero at a freezing time
$\left\langle T\right\rangle $, stopping the dynamics, with a finite
fraction $n_{u,a}$ of unsatisfied living cells which becomes smaller
as $\rho_{d}^{\text{(mf)}}$ is approached from below (see Figs. \ref{fig:Ninf1vsf_low}
and \ref{fig:Ninf0vst_mf}), while the freezing time becomes larger.
At the critical value $\rho_{d}^{\text{(mf)}}$, as shown in Fig.
\ref{fig:Ninf0vst_mf}, the mean-field freezing time tends to a finite
value $\left\langle T\right\rangle ^{\text{(mf)}}\simeq3.23148$,
at which $n_{u,d}$ and its time derivative both become zero, right
before $n_{u,d}$ rises again, approaching a value $3.4\times10^{-4}$
as $t\rightarrow\infty$. For larger values of $\rho_{d}$ there are
no zeros for any $\phi_{s,k}\left(t\right)$, meaning that the freezing
time is infinite, a signature of the active phase. Finally, as shown
in the left inset of Fig. \ref{fig:Ninf1vsf_low}, the fraction of
unsatisfied living cells $n_{u,a}$ as $\rho_{d}$ is increased towards
$\rho_{d}^{\text{(mf)}}$ approaches a nonzero value $n_{u,a}^{\text{(mf)}}\simeq0.263120$,
which differs only slightly from the mean-field result $n_{u,a}\left(t\rightarrow\infty\right)\simeq0.258111$
obtained throughout the active phase, as seen in the blue dashed curve
in Fig. \ref{fig:Ninf0vst_mf}. 

Therefore, in agreement with simulations, the mean-field approximation
predicts a discontinuous transition between a small-$\rho_{d}$ inactive
phase and an intermediate-$\rho_{d}$ active phase. As shown in Fig.
\ref{fig:Ninf1vsf_low}, there is quite good quantitative agreement
between simulations and mean-field theory for densities $\rho_{d}\lesssim0.2$,
regarding the fraction of unsatisfied living cells upon freezing,
but also (not shown) the average freezing time. Notice also the qualitative
agreement, in the neighborhood of the transition, between the behaviors
of the curves for $n_{u,d}\left(t\right)$ obtained from the mean-field
calculation (continuous red and brown curves in Fig. \ref{fig:Ninf0vst_mf})
and the large-$L$ simulation results (extrapolation of the curves
in Fig. \ref{fig:Ninf1vsf_low}(b)).

The freezing time remains infinite for intermediate densities of dead
cells, $\rho_{d}^{\text{(mf)}}<\rho_{d}<\rho_{d}^{\text{(mf2)}}\simeq0.878211$,
but becomes finite again for higher densities, $\rho_{d}^{\text{(mf2)}}<\rho_{d}<1$,
signaling the onset of a second, large-$\rho_{d}$ inactive phase,
although simulations predict a critical-like behavior for the whole
region $\rho_{d}^{(2)}<\rho_{d}<1$. This discrepancy between simulation
and mean-field theory comes from the fact that, as discussed towards
the end of Sec. \ref{sec:simulations}, at higher densities most exchanges
on the square lattice involve isolated unsatisfied living cells with
8 dead neighbors, virtually all of which remain unsatisfied after
the switching, so that new unsatisfied dead cells mostly appear in
the neighborhood of the rare unsatisfied dead cells generated by the
initial condition. Such correlations between dead cells cannot be
captured by our mean-field treatment.

\section{Conclusions}

\label{sec:discussion}We showed, through simulations and a mean-field
calculation, that a conservative version of the Game of Life exhibits
two nonequilibrium phase transitions separating an active phase from
two distinct inactive phases as the density $\rho_{d}$ of dead cells
is increased. The long-time fractions $n_{u,a}$ and $n_{u,d}$ of
unsatisfied living and dead cells are both nonzero in the active phase,
while, in the thermodynamic limit, the dynamics in the small-$\rho_{d}$
(large-$\rho_{d}$) inactive phase is interrupted when $n_{u,d}$
($n_{u,a}$) becomes zero. The transition between the small-$\rho_{d}$
inactive phase and the active phase is discontinuous, according to
both simulations and mean-field theory. On the other hand, simulations
show that the transition between the active phase and the large-$\rho_{d}$
phase is continuous, with a set of critical exponents that, to the
best of our knowledge, does not correspond to any of the known universality
classes for multicomponent nonequilibrium systems \citep{Odor2008}.

Improvements to the mean-field approximation can be achieved by analyzing
all $2^{8}$ possible local configurations of neighbors surrounding
a given cell in order to identify the restrictions imposed by the
fact that some of the neighbors of the central cell are also mutual
neighbors. This requires dealing not with the fractions of cells having
a given number of living cells in their neighborhood, but with the
fractions of cells with a specific configuration of neighbors. Thus,
instead of a set of 16 differential equations, one ends up with a
set of $2^{9}$ differential equations to write and solve. We leave
this improvement for future investigations. The simple mean-field
calculation described in this paper can nevertheless be used to study
phase transitions in other conservative models, such as the extensions
of Schelling's model recently investigated by the present authors
\citep{Vieira2020}. 

Finally, an interesting question arises from the fact that other nonequilibrium
models, such as the contact process, can be equivalently defined either
(i) in terms of fixed transition rates, with no conserved quantities,
the densities of each type of cell being determined by the dynamics,
or (ii) by fixing the densities of each type of cell, with the average
value of the transition rate determined by the dynamics \citep{tome2001}.
One might wonder whether there is a nonconservative version of the
Game of Life discussed here, but defined in terms of fixed transition
rates, and how that would be related to the original Game of Life.
The fact that the dynamical rules are non-Abelian --- in other words,
that the order in which movements are performed does affect the resulting
configurations --- suggests that such relation, if it indeed exists,
would not be simple. This is also the reason why we resist the temptation
to associate the large-$\rho_{d}$ inactive phase identified in our
work with the quasicritical behavior of the original automaton \citep{Alstrom1994,Hemmingsson1996,Blok1997,Reia2014}.\bigskip{}

\begin{acknowledgments}
This work was supported by the Brazilian agencies FUNCAP, CAPES, CNPq,
INCT-SC, NAP-FCx, INCT-FCx and FAPESP. EG acknowledges financial support
from Fondecyt-Anid 1200006.
\end{acknowledgments}

\bibliography{agents}

\end{document}